\documentclass[]{spie}  
 \usepackage{pbox}
\usepackage{multirow}
\usepackage{amsmath,amsfonts,amssymb}
\usepackage{graphicx}
\usepackage{floatrow}
\usepackage{aas_macros}
\usepackage[export]{adjustbox}
\graphicspath{{./figures/}} 
\usepackage{sidecap}
\usepackage[colorlinks=true, allcolors=blue]{hyperref}
\usepackage{enumerate}
     \usepackage[flushleft]{threeparttable}
     \usepackage{array}
\usepackage{booktabs}
\usepackage [english]{babel}
\usepackage [autostyle, english = american]{csquotes}
\MakeOuterQuote{"}

\title{Panoramic SETI: Program Update and High-Energy Astrophysics Applications}
\author[a]{J\'er\^ome Maire}
\author[a,b]{Shelley A. Wright}
\author[c]{Jamie Holder}
\author[d]{David Anderson}
\author[e]{Wystan Benbow}
\author[a]{Aaron Brown}
\author[a,b]{Maren Cosens}
\author[c]{Gregory Foote}
\author[e]{William F. Hanlon}
\author[f]{Olivier Hervet}
\author[g]{Paul Horowitz}
\author[h]{Andrew W. Howard}
\author[d]{Ryan Lee}
\author[d,i]{Wei Liu}
\author[j]{Rick Raffanti}
\author[d,i]{Nicolas Rault-Wang}
\author[k]{Remington P. S. Stone}
\author[d,i]{Dan Werthimer}
\author[a,b]{James Wiley}
\author[f]{David A. Williams}
\affil[a]{Center for Astrophysics \& Space Sciences, University of California San Diego, CA, USA}
\affil[b]{Department of Physics, University of California San Diego, CA, USA}
\affil[c]{Department of Physics and Astronomy, University of Delaware, DE, USA.}
\affil[d]{Space Sciences Laboratory, University of California Berkeley, CA, USA}
\affil[e]{Center for Astrophysics $|$ Harvard \& Smithsonian, Cambridge, MA, USA}
\affil[f]{Santa Cruz Institute for Particle Physics and Department of Physics, University of California, Santa Cruz, CA, USA}
\affil[g]{Department of Physics, Harvard University, Cambridge, MA, USA}
\affil[h]{Astronomy Department, California Institute of Technology, Pasadena, CA, USA}
\affil[i]{Department of Astronomy, University of California Berkeley, CA, USA}
\affil[j]{Techne Instruments, Oakland, CA, USA}
\affil[k]{University of California Observatories, Lick Observatory, CA, USA}

\authorinfo{Further author information: (Send correspondence to J.M.)\\J.M.: E-mail: jmaire@ucsd.edu}

\begin{document} 
\maketitle

\begin{abstract}
Optical SETI (Search for Extraterrestrial Intelligence) instruments that can explore the very fast time domain, especially with large sky coverage, offer an opportunity for new discoveries that can complement multimessenger and time domain astrophysics. The Panoramic SETI experiment (PANOSETI) aims to observe optical transients with nanosecond to second duration over a wide field-of-view ($\thicksim$2,500\,sq.deg.) by using two assemblies of tens of telescopes to reject spurious signals by coincidence detection.
Three PANOSETI telescopes, connected to a White Rabbit timing network used to synchronize clocks at the nanosecond level, have been deployed at Lick Observatory on two sites separated by a distance of 677 meters to distinguish nearby light sources (such as Cherenkov light from particle showers in the Earth’s atmosphere) from astrophysical sources at large distances.
In parallel to this deployment, we present results obtained during four nights of simultaneous observations with the four 12-meter VERITAS gamma-ray telescopes and two PANOSETI telescopes at the Fred Lawrence Whipple Observatory.
We report PANOSETI’s first detection of astrophysical gamma rays, comprising three events with energies in the range between $\thicksim$15 TeV and $\thicksim$50 TeV. These were emitted by the Crab Nebula, and identified as gamma rays using joint VERITAS observations.

\end{abstract}


\section{INTRODUCTION}

Astronomical instruments with precision time resolution can be employed 
to search for technosignatures by means of detecting nano- to milli-second light pulses that could be emitted, for instance, for the purpose of interstellar communications or energy transfer. Numerous programs have been conducted to search for technosignatures using optical  wavelengths \cite{Werthimer2001, Horowitz2001, Reines2002, Howard2004, Stone2005, Abeysekara2016, Tellis2017} including near-infrared\cite{Maire2019}. With 0.32\,sq.deg.\@ of instantaneous field-of-view, the first optical SETI all-sky surveys\cite{Howard2000, Howard2007} used a transit observing approach to cover the sky in 150 clear nights. To efficiently and constantly survey the entire sky, however, groups of single-aperture telescopes capable  of  collecting light instantaneously from  different  parts  of  the  sky  are  still required.

The Panoramic SETI  experiment (PANOSETI\cite{Maire2018,Wright2018,Cosens2018,Wright2019}) is an all-sky observatory project aiming to detect transients that will cover a wide range of timescales in a search for nanosecond to second pulsed light signals across all optical wavelengths. Each part of the sky is observed simultaneously from two locations for direct detection and confirmation of transients. Based upon two assemblies of twenty-four 0.46-m Fresnel-lens telescopes equipped with fast, low-noise silicon photo-multipliers operating in the 0.32--0.85\,$\mu$m spectral range, the PANOSETI instrument can detect flashes of light that could have been sent from kiloparsec distances and beamed toward our direction. For instance, a pulse of light emitted by a laser delivering 20\,ns shaped pulses of megajoule energies and collimated with a 10-m telescope located at 1\,kpc away from us would be orders of magnitude brighter than the entire broadband visible stellar background from our perspective\cite{Howard2004}. 

The small aperture, wide field-of-view, and low cost of the PANOSETI telescopes also make them potentially well-suited for gamma-ray astronomy at the highest energies. When a high-energy gamma-ray photon or cosmic-ray particle enters the Earth's atmosphere, it initiates a particle cascade which can be detected via the pulse of blue Cherenkov radiation it emits. Ground-based gamma-ray telescopes exploit this effect, using large aperture ($>$ 10\,m) mirrors and fast photo-detector cameras to record Cherenkov images of the showers. Subsequent analysis of these images allows determination of the nature of the primary (photon or particle), its arrival direction and its energy. The energy range covered by this technique is typically from around 100 GeV to tens of TeV, reaching 100 TeV only for the brightest sources. Hundreds of astrophysical gamma-ray sources, both Galactic and extragalactic, have been identified using this technique\footnote{\url{http://tevcat.uchicago.edu/}}. At higher energies, direct detection of the shower particles is used; recent results from the LHAASO collaboration have revealed the existence of gamma-ray emitters extending up to PeV energies\cite{Cao2021a}. The main benefit of this particle detection approach is that it provides the large effective area ($>$1\, km$^2$) which is required to measure the extremely low flux of gamma rays at the highest energy. Cherenkov telescopes can operate in the same energy range, but are typically non-imaging, to reduce unit cost (e.g., TAIGA-HiScore \cite{2020NIMPA.95862113B}). The unit cost of PANOSETI telescopes is approximately two orders of magnitude less than a large aperture imaging Cherenkov telescope, such as those that make up the VERITAS array. A large array of low-cost PANOSETI telescopes, each equipped with a 1,000 pixel camera, would permit applying the benefits of the imaging technique (excellent angular resolution, energy resolution and background discrimination) to the PeV regime. 

We describe in Sect.\ref{sec:program} the recent deployment of a third PANOSETI telescope at Lick Observatory  providing a baseline separation of 677-m. We report in Sect.\ref{sec:veritas} results obtained from joint PANOSETI and VERITAS observations at the Whipple Observatory in November 2021.

\section{PANOSETI: Program Update}\label{sec:program}

The PANOSETI experiment aims to observe 2,350 square degrees instantaneously by making use of multiple large field-of-view telescopes. PANOSETI is currently in its final design phase, and at final production two dedicated observatories will house 24 telescopes per site. 
Each part of the sky is observed simultaneously from two locations for direct detection and confirmation of optical transients.

Each telescope is equipped with a 0.46-m f$/$1.32 Fresnel-lens which focuses the light onto a  32x32-pixel photon-counting detector subdivided into 16 adjacent 8x8-pixel Multi-Pixel Photon Counter (Hamamatsu S13361-3050AE-08) detector arrays  operating in the 0.32\,-\,0.85\,$\mu$m spectral range. These silicon photomultipliers (SiPMs) are comprised of Geiger-mode-operated avalanche photodiodes highly linear in pulse intensity, with a high internal gain to enable single-photon detection while featuring low dark count ($<$1\, Mcps per SiPM), high photon detection efficiency in the visible (45\%), and nanosecond timing resolution. 
Each PANOSETI detector contains four custom readout boards that make use of a 64-bit Application-Specific Integrated Circuit (ASIC) capable of pulse shaping and trigger detection of individual pulses. Each detector board reads four 8x8 pixel SiPM arrays that feed four ASICs, amplifying the signal and delivering a per-pixel trigger signal to a Field Programmable Gate Array (FPGA). A 1\,Gb per second fiber connection provides data communications to a central multi-telescope 10\,Gb network switch connected to a central computer.

For timing synchronicity between telescopes, 
PANOSETI makes use of a White Rabbit\cite{Moreira2009, Liu2020} precision time distribution system that provides 1\,ns time-stamping for each triggered event. The PANOSETI White Rabbit system is connected to a GPS receiver for absolute timing synchronicity.

To detect optical pulse widths ranging from nanoseconds to seconds, each telescope's custom readout electronics system  is configured to support two different operational modes that can be run simultaneously\cite{Liu2020}.
 The Pulse Height mode starts data acquisition when any  pixel signal exceeds a programmable threshold, triggering the readout of  the entire quadrant board. These events are then cross-correlated in time with the other telescopes to check for any other triggered events. This Pulse Height mode has been standard for previous optical SETI experiments \cite{Wright2001, Werthimer2001, Horowitz2001, Howard2004, Wright2014} and is optimized for detecting pulse widths shorter than $\thicksim$100\,ns. Concurrently, the Continuous Imaging Mode employs counters on every pixel that count over-threshold events to produce images at a programmable frame rate. This mode is used for detecting transients above 20 microsecond duration. Over-threshold events for each pixel are counted by the FPGA and all pixel counters are sent over ethernet to the central computer system, along with an accurate time stamp for each frame. 
 
Our team is conducting laboratory and on-sky verification of our prototype telescopes and focal plane electronics. To verify our White Rabbit timing system and false alarm rates, we first installed two prototype units at Lick Observatory at a single site to measure on-sky performance. 

\begin{figure}[ht]
\includegraphics[width=16cm]{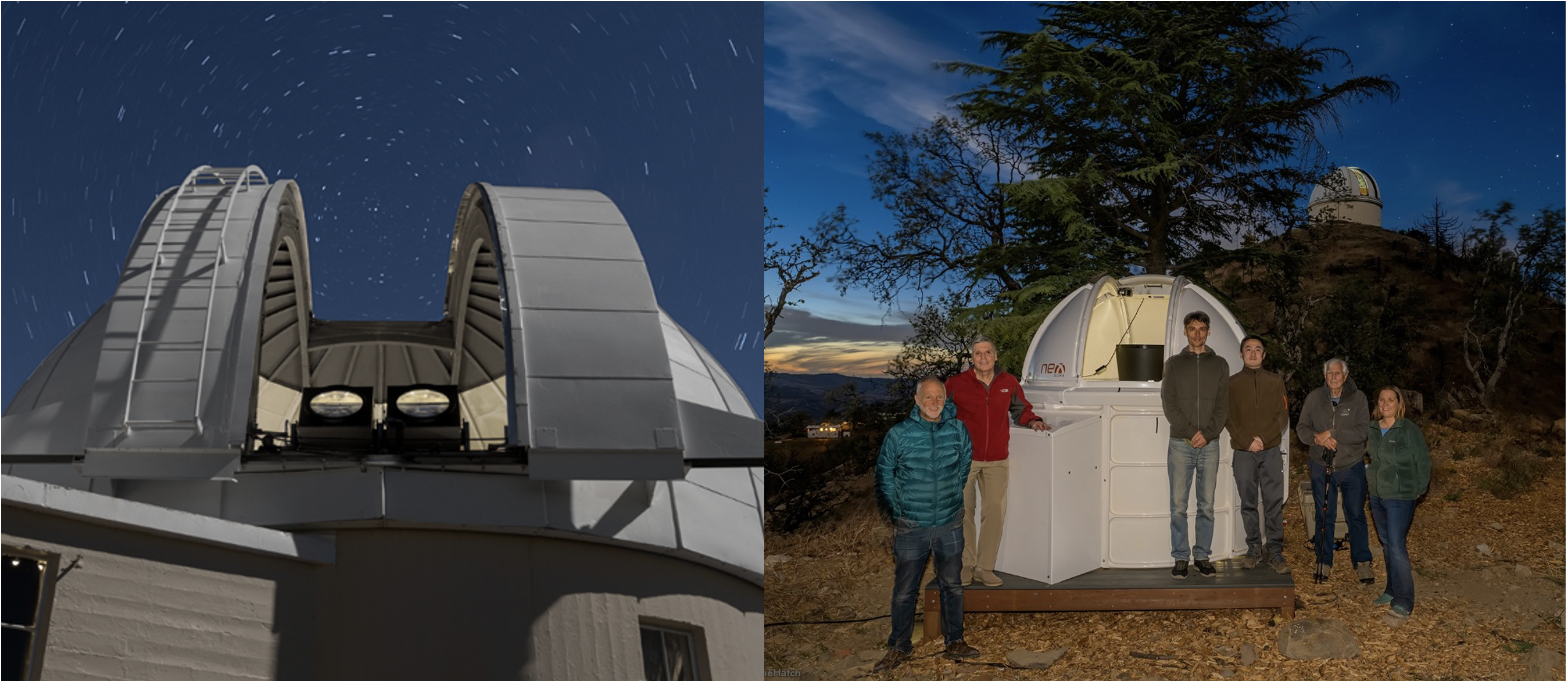}
    \caption{LEFT: Two PANOSETI telescopes were deployed inside the Astrograph dome at Lick Observatory. RIGHT: The deployment of a third telescope at a distance of 677\,m from the Astrograph confirmed the benefit of using a long baseline to identify and characterize transient phenomena (Left and right photos: courtesy Laurie Hatch Photography). 
}
    \label{fig:panolick}
\end{figure}

During prototyping efforts, the deployment of pairs of telescopes has proven the benefit of using long baselines to minimize the number of false alarms by providing enough parallax to easily distinguish nearby transient light sources (such as Cherenkov showers) from astrophysical sources at large distances \cite{Maire2020}. Therefore, a third telescope  was deployed at Lick Observatory in June 2022 at a distance of 677 meters from the first site (Fig.\ref{fig:panolick}). It uses a 2.2-m diameter remote-controlled dome and is fiber-connected to the first site for data transfer, remote operations, and time synchronization.

\section{PANOSETI Coordinated Observations with VERITAS}\label{sec:veritas}
\subsection{Experimental setup}\label{subsec:setup}
\begin{figure}[ht]
\includegraphics[width=11cm]{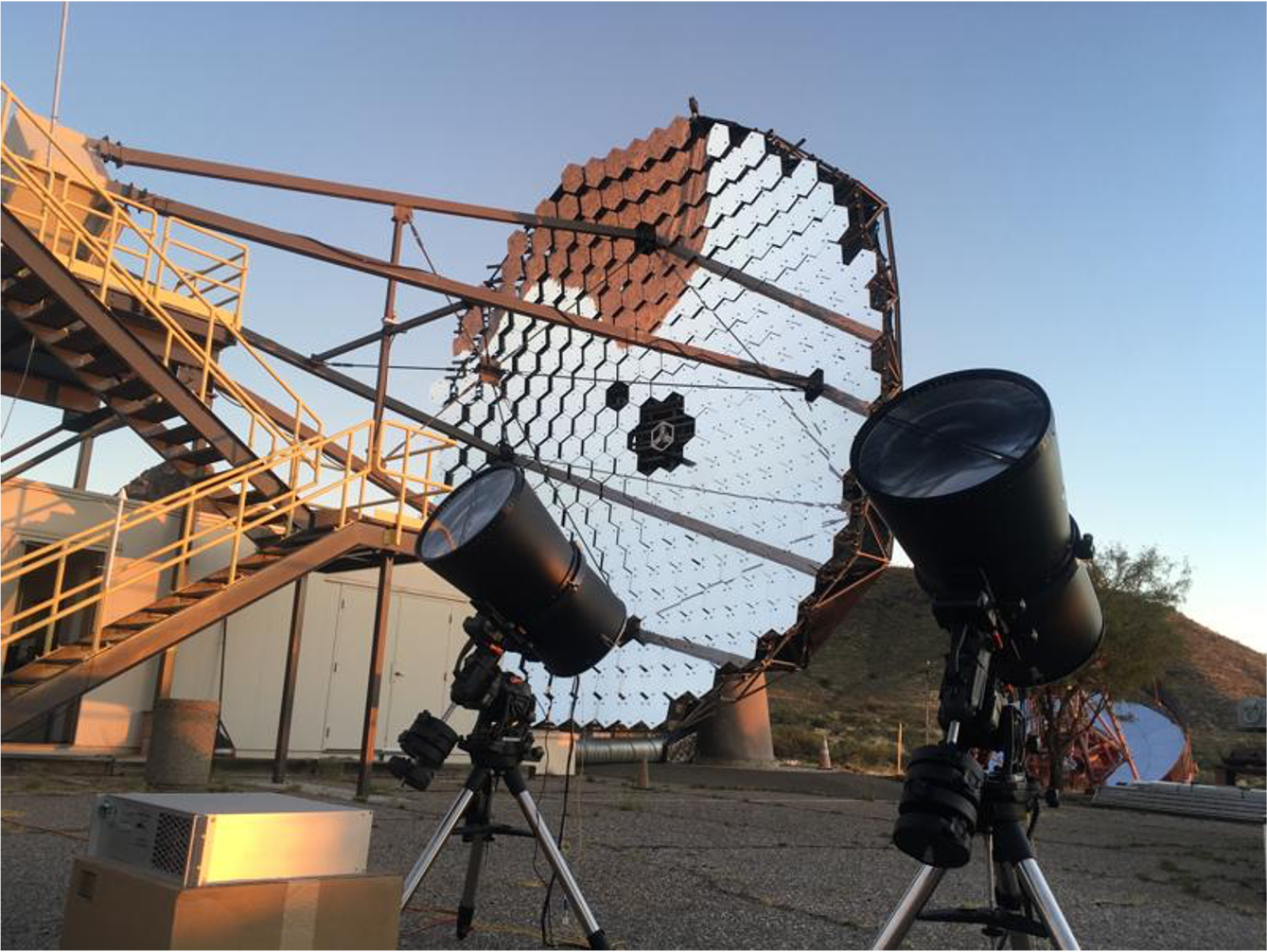}
    \caption{Pictured here are two PANOSETI 0.5-m telescopes with one of the 12-m VERITAS telescopes (T4) in the background. Joint observations were performed in November 2021.}
    \label{fig:pano}
\end{figure}

The PANOSETI team performed four nights of simultaneous observations with the four 12-meter VERITAS gamma-ray telescopes \cite{Weekes2002} at the Fred Lawrence Whipple Observatory on November 4-7 2021. Coordinated observations of CALIPSO (LIDAR satellite), the Crab Nebula, and other potential sources of gamma rays were performed. 
During the first two nights of coordinated observations, the two PANOSETI telescopes were located beside the VERITAS T4 telescope (Fig.\ref{fig:pano}). For the last two nights, one of the PANOSETI telescopes was moved beside the VERITAS T2 telescope  at a distance of 127 meters from the other PANOSETI telescope. Two long fibers were temporarily deployed for data communications between the telescopes, the host computer and the White Rabbit switch.

During all of the coordinated observations, the PANOSETI telescopes were aligned to the same celestial coordinates targeted by the four VERITAS telescopes and made use of Celestron CGX equatorial mounts for tracking. The PANOSETI telescope backends were also aligned in position angle.

\subsection{Data \& Analysis}\label{subsec:analysis}
During joint observations, the PANOSETI detectors were set in Pulse Height mode to detect and record events above a minimum threshold of 11.5 photo-electrons in a single pixel. 
All PANOSETI triggered events recorded during the Crab Nebula observations were cross-correlated in time with the VERITAS detected events to search for matched events.  The same procedure was used to search for events coincident in time between PANOSETI telescopes that occurred within the same 100\,ns time window, after correcting for time delays due to the separation and pointing directions of the telescopes.

\begin{figure}[ht]
\includegraphics[width=7cm]{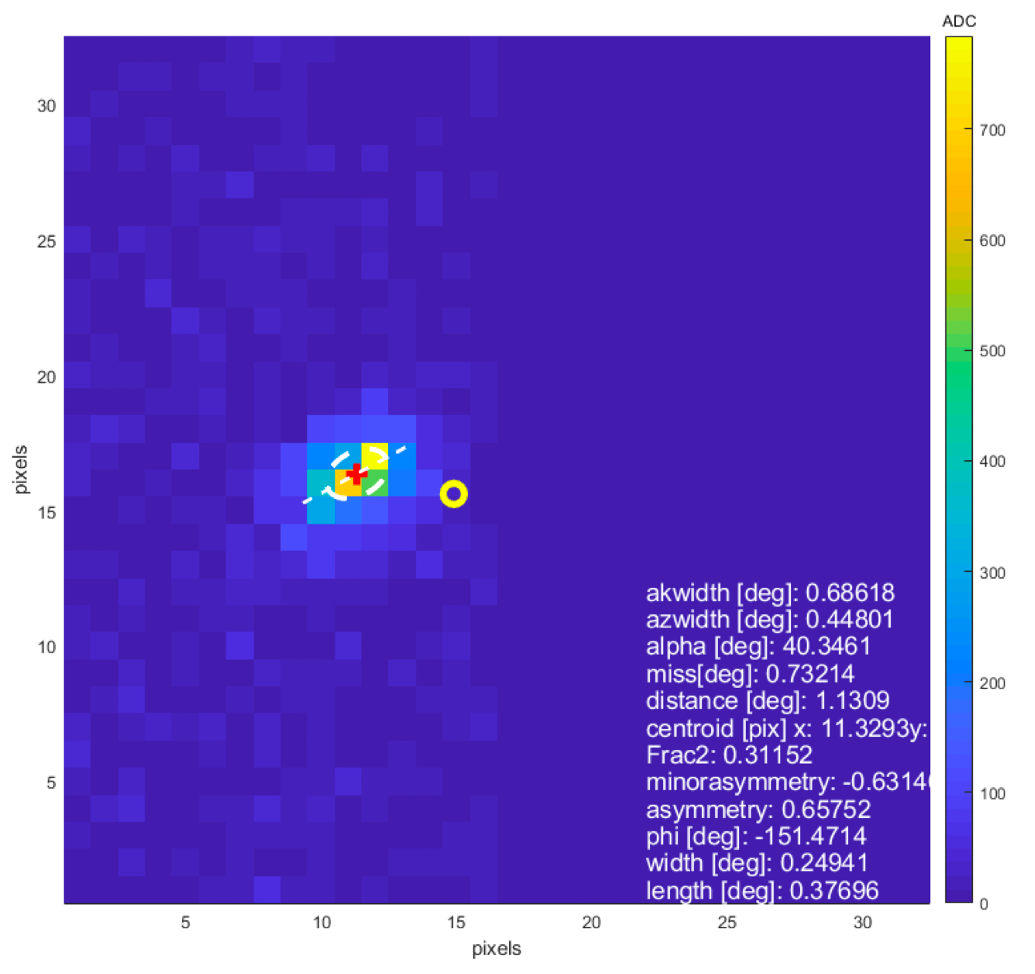}
 \includegraphics[width=6.5cm]{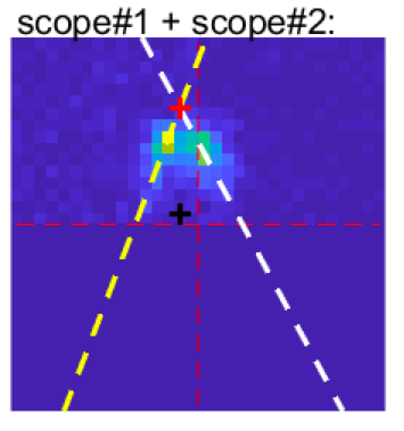}
    \caption{Examples of hadron-induced showers imaged with PANOSETI along with Hillas parameters (left panel). The intersection of shower main axes measured on the two telescopes determines the arrival direction (right panel).
}
    \label{fig:panohillas}
\end{figure}

Each PANOSETI shower image is parameterized using the Hillas method \cite{Fegan1997} to characterize the morphology of the shower (such as the direction of shower main axis, the length and the width of the apparent shower, and its asymmetry, as illustrated on Figure \ref{fig:panohillas}-left). A direct comparison of Hillas parameter values  with those obtained with VERITAS data is then possible. A subsequent step of intersecting the shower main axes measured by the two separated PANOSETI telescopes allows us to determine the arrival direction of the hadron or gamma ray that induced the Cherenkov shower (Figure \ref{fig:panohillas}-right), as performed routinely with data from the four VERITAS telescopes\cite{Fegan1997}.
The PANOSETI 127-meter-long baseline used during the Crab Nebula observations introduces a small but measurable parallax angle for flashes of light generated in Earth's atmosphere ($\thicksim$0.2$^\circ$ to $\thicksim$0.8$^\circ$, i.e., $\thicksim$0.7 to $\thicksim$2.6 pixels for a shower maximum at 10\,km above sea level and above 30$^\circ$ elevation).

\subsection{Results}\label{subsec:res}

\subsubsection{Timing synchronicity between VERITAS and PANOSETI
}\label{subsubsec:calipso}

Joint observations of the NASA/CNES Cloud-Aerosol Lidar and Infrared Pathfinder Satellite (CALIPSO), which sends repeated 20\,ns optical pulses at a frequency of 20.16Hz, were performed at the Fred Lawrence Whipple Observatory on November 6 2021. CALIPSO was observed during culmination at 80.6$^\circ$ elevation. VERITAS T4 telescope detected $\thicksim$75 pulses, with an average pulse intensity of $\thicksim$4000 photo-electrons. PANOSETI detected 180 pulses
with peak intensities between  $\thicksim$20 and 50 photo-electrons, depending on how the satellite was centered on the pixels.

\begin{figure}[ht]
\includegraphics[width=11cm]{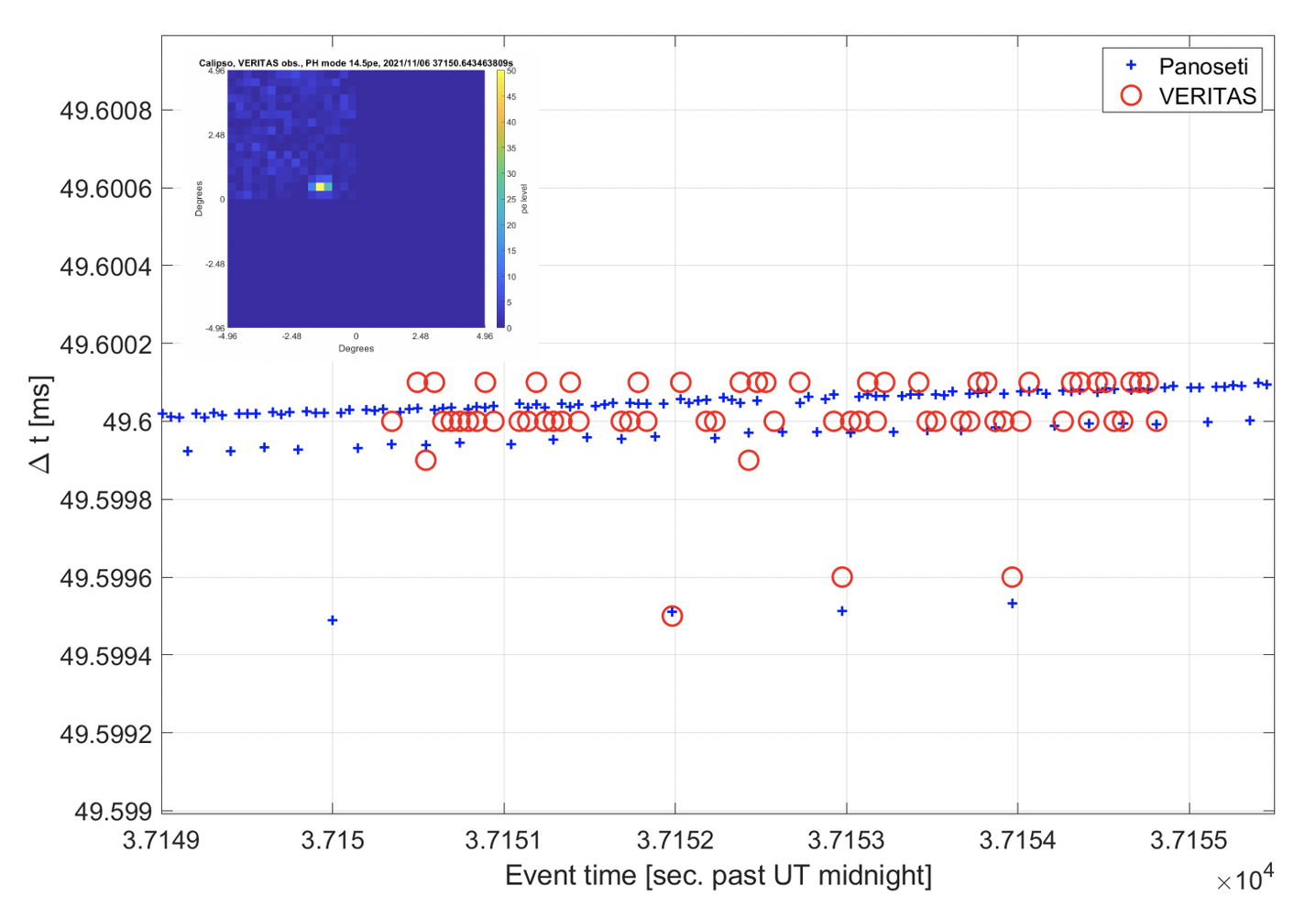}
    \caption{(top-left) CALIPSO optical pulse as seen by one of the PANOSETI telescopes at the Whipple Observatory on November 6 2021. The figure shows the difference in time between successive pulses obtained with one of the PANOSETI telescopes and the closest VERITAS telescope (T4), showing sub-microsecond synchronicity between the VERITAS and PANOSETI telescopes. PANOSETI was some 120km distant from CALIPSO's 70m footrpint: glints from the latter's aperture were adequate for 100\% detection.
}
    \label{fig:panoveritastiming}
\end{figure}

Figure \ref{fig:panoveritastiming} shows the difference in time between successive timestamps obtained during joint observations of CALIPSO. These simultaneous measurements of the CALIPSO pulses confirmed  the sub-microsecond synchronicity between PANOSETI and VERITAS telescopes.

\subsubsection{PANOSETI gamma-ray detections}\label{subsubsec:gamma}
The night of November 7 2021 was dedicated to joint observations of the Crab Nebula, during which one of the PANOSETI telescopes was deployed beside the VERITAS T4 telescope. The PANOSETI telescope hardware was fiber-connected for data transfer and timing synchronicity to the other PANOSETI telescope, itself deployed beside the VERITAS T2 telescope. 

\begin{figure}[ht]
\includegraphics[width=13cm]{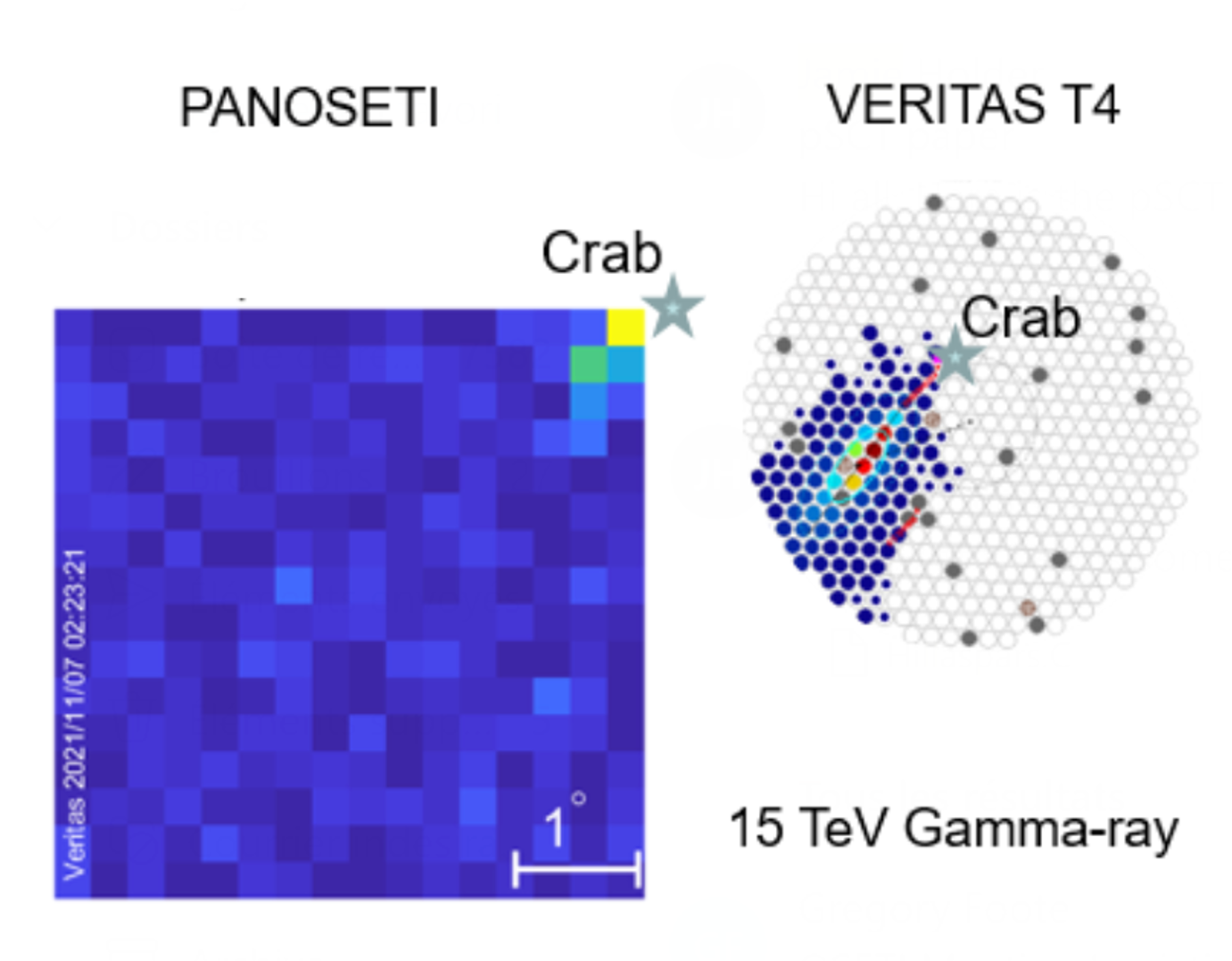}
    \caption{One of the three Cherenkov showers that were induced by gamma rays emitted from the Crab Nebula, imaged by one of the PANOSETI  detector quadrants (left) and VERITAS T4 telescope (right). This event was identified by the four VERITAS telescopes to be a gamma ray with an energy of 15\,TeV. Detected within the same microsecond, the length, width and orientation of the shower measured by the Hillas parameters are in excellent agreement between the two instruments.
}
    \label{fig:panoveritasgammaray}
\end{figure}

During 6.2 hours of continuous observations, 10,540 Cherenkov showers (largely cosmic-ray background) observed with PANOSETI were confirmed by VERITAS.  Among these, 3,122 showers were detected by the two PANOSETI telescopes set to detect events above 11.5 photo-electrons. Three of these showers were identified by the four VERITAS telescopes as being very likely due to gamma rays emitted by the Crab Nebula (Fig. \ref{fig:panoveritasgammaray}). The reconstructed energy of the primary gamma rays obtained from the VERITAS data analysis was 15, 36 and 51\,TeV.

\section{Summary}\label{sec:Sum}

 We have demonstrated that PANOSETI  has the potential to explore the extreme Universe through the detection of ultra-high-energy gamma rays.  PANOSETI is capable of detecting gamma rays (and cosmic rays) with energies greater than $\thicksim$10 TeV. PANOSETI’s first detection of gamma rays emitted from the Crab Nebula was confirmed by joint VERITAS observations. 

With its  677-m long-baseline prototype telescopes recently deployed at Lick Observatory, PANOSETI is entering its final design phase with on-sky characterizations. At final production, two dedicated observatories will house 24 telescopes per site. New pairs of telescopes will be added progressively to increase the size of its total instantaneous field-of-view,  for an all-sky search of optical technosignatures and  astrophysical transients.

\acknowledgments 
We would like to acknowledge the Lick Observatory staff and engineers for their help in the installation of the PANOSETI telescopes into the Astrograph dome and Barnard sites. We also thank the VERITAS Collaboration and the technical support staff at the Fred Lawrence Whipple Observatory for their cooperation in obtaining joint observations and for the use of their data. We thank the Bloomfield Family Foundation for supporting SETI research at UC San Diego in the CASS Optical and Infrared Laboratory. We also thank Franklin Antonio for valuable comments and support. Harvard SETI was supported by The Planetary Society and The Bosack/Kruger Charitable Foundation. UC Berkeley's SETI efforts involved with PANOSETI are supported by NSF grant 1407804, the Breakthrough Prize Foundation, and the Marilyn and Watson Alberts SETI Chair fund. 


\bibliography{main2} 

\begin{thebibliography}{10}

\bibitem{Werthimer2001}
{Werthimer}, D., {Anderson}, D., {Bowyer}, C.~S., {Cobb}, J., {Heien}, E.,
  {Korpela}, E.~J., {Lampton}, M.~L., {Lebofsky}, M., {Marcy}, G.~W.,
  {McGarry}, M., and {Treffers}, D., ``{Berkeley radio and optical SETI
  programs: SETI@home, SERENDIP, and SEVENDIP},'' in [{\em The Search for
  Extraterrestrial Intelligence (SETI) in the Optical Spectrum
  III}{\nolinebreak\hspace{0.1em}]},  {Kingsley}, S.~A. and {Bhathal}, R.,
  eds., {\em Proc. SPIE, Vol.4273} {\bf 4273},  104--109 (Aug. 2001).

\bibitem{Horowitz2001}
{Horowitz}, P., {Coldwell}, C.~M., {Howard}, A.~B., {Latham}, D.~W.,
  {Stefanik}, R., {Wolff}, J., and {Zajac}, J.~M., ``{Targeted and all-sky
  search for nanosecond optical pulses at Harvard-Smithsonian},'' in [{\em The
  Search for Extraterrestrial Intelligence (SETI) in the Optical Spectrum
  III}{\nolinebreak\hspace{0.1em}]},  {Kingsley}, S.~A. and {Bhathal}, R.,
  eds., {\em Society of Photo-Optical Instrumentation Engineers (SPIE)
  Conference Series} {\bf 4273},  119--127 (Aug. 2001).

\bibitem{Reines2002}
{Reines}, A.~E. and {Marcy}, G.~W., ``{Optical Search for Extraterrestrial
  Intelligence: A Spectroscopic Search for Laser Emission from Nearby Stars},''
  {\em PASP}~{\bf 114},  416--426 (Apr. 2002).

\bibitem{Howard2004}
{Howard}, A.~W., {Horowitz}, P., {Wilkinson}, D.~T., {Coldwell}, C.~M.,
  {Groth}, E.~J., {Jarosik}, N., {Latham}, D.~W., {Stefanik}, R.~P., {Willman},
  Jr., A.~J., {Wolff}, J., and {Zajac}, J.~M., ``{Search for Nanosecond Optical
  Pulses from Nearby Solar-Type Stars},'' {\em ApJ}~{\bf 613},  1270--1284
  (Oct. 2004).

\bibitem{Stone2005}
{Stone}, R.~P.~S., {Wright}, S.~A., {Drake}, F., {Mu{\~n}oz}, M., {Treffers},
  R., and {Werthimer}, D., ``{Lick Observatory Optical SETI: Targeted Search
  and New Directions},'' {\em Astrobiology}~{\bf 5},  604--611 (Oct. 2005).

\bibitem{Abeysekara2016}
{Abeysekara}, A.~U., {Archambault}, S., {Archer}, A., {Benbow}, W., {Bird}, R.,
  {Buchovecky}, M., {Buckley}, J.~H., {Byrum}, K., {Cardenzana}, J.~V.,
  {Cerruti}, M., {Chen}, X., {Christiansen}, J.~L., {Ciupik}, L., {Cui}, W.,
  {Dickinson}, H.~J., {Eisch}, J.~D., {Errando}, M., {Falcone}, A., {Fegan},
  D.~J., {Feng}, Q., {Finley}, J.~P., {Fleischhack}, H., {Fortin}, P.,
  {Fortson}, L., {Furniss}, A., {Gillanders}, G.~H., {Griffin}, S., {Grube},
  J., {Gyuk}, G., {H{\"u}tten}, M., {H{\r{a}}kansson}, N., {Hanna}, D.,
  {Holder}, J., {Humensky}, T.~B., {Johnson}, C.~A., {Kaaret}, P., {Kar}, P.,
  {Kelley-Hoskins}, N., {Kertzman}, M., {Kieda}, D., {Krause}, M., {Krennrich},
  F., {Kumar}, S., {Lang}, M.~J., {Lin}, T.~T.~Y., {Maier}, G., {McArthur}, S.,
  {McCann}, A., {Meagher}, K., {Moriarty}, P., {Mukherjee}, R., {Nieto}, D.,
  {O'Brien}, S., {O'Faol{\'a}in de Bhr{\'o}ithe}, A., {Ong}, R.~A., {Otte},
  A.~N., {Park}, N., {Perkins}, J.~S., {Petrashyk}, A., {Pohl}, M., {Popkow},
  A., {Pueschel}, E., {Quinn}, J., {Ragan}, K., {Ratliff}, G., {Reynolds},
  P.~T., {Richards}, G.~T., {Roache}, E., {Santander}, M., {Sembroski}, G.~H.,
  {Shahinyan}, K., {Staszak}, D., {Telezhinsky}, I., {Tucci}, J.~V., {Tyler},
  J., {Vincent}, S., {Wakely}, S.~P., {Weiner}, O.~M., {Weinstein}, A.,
  {Williams}, D.~A., and {Zitzer}, B., ``{A Search for Brief Optical Flashes
  Associated with the SETI Target KIC 8462852},'' {\em \apjl}~{\bf 818},  L33
  (Feb. 2016).

\bibitem{Tellis2017}
{Tellis}, N.~K. and {Marcy}, G.~W., ``{A Search for Laser Emission with
  Megawatt Thresholds from 5600 FGKM Stars},'' {\em AJ}~{\bf 153},  251 (June
  2017).

\bibitem{Maire2019}
{Maire}, J., {Wright}, S.~A., {Barrett}, C.~T., {Dexter}, M.~R., {Dorval}, P.,
  {Duenas}, A., {Drake}, F.~D., {Hultgren}, C., {Isaacson}, H., {Marcy}, G.~W.,
  {Meyer}, E., {Ramos}, J.~R., {Shirman}, N., {Siemion}, A., {Stone}, R. P.~S.,
  {Tallis}, M., {Tellis}, N.~K., {Treffers}, R.~R., and {Werthimer}, D.,
  ``{Search for Nanosecond Near-infrared Transients around 1280 Celestial
  Objects},'' {\em Astronomical Journal}~{\bf 158},  203 (Nov. 2019).

\bibitem{Howard2000}
{Howard}, A.~W., {Horowitz}, P., and {Coldwell}, C.~M., ``An all-sky optical
  seti survey,'' in [{\em Proceedings of the 51st IAF Congress in Rio de
  Janeiro, Brazil, Acta Astronautica}{\nolinebreak\hspace{0.1em}]},  (2000).

\bibitem{Howard2007}
Howard, A., Horowitz, P., Mead, C., Sreetharan, P., Gallicchio, J., Howard, S.,
  Coldwell, C., Zajac, J., and Sliski, A., ``{Initial results from Harvard
  all-sky optical SETI},'' {\em Acta Astronautica}~{\bf 61},  78--87 (June
  2007).

\bibitem{Maire2018}
{Maire}, J., {Wright}, S.~A., {Cosens}, M., {Antonio}, F.~P., {Aronson}, M.~L.,
  {Chaim-Weismann}, S.~A., {Drake}, F.~D., {Horowitz}, P., {Howard}, A.~W.,
  {Marcy}, G.~W., {Raffanti}, R., {Siemion}, A. P.~V., {Stone}, R. P.~S.,
  {Treffers}, R.~R., {Uttamchandani}, A., and {Werthimer}, D., ``{Panoramic
  optical and near-infrared SETI instrument: optical and structural design
  concepts},'' in [{\em Ground-based and Airborne Instrumentation for Astronomy
  VII}{\nolinebreak\hspace{0.1em}]},  {Evans}, C.~J., {Simard}, L., and
  {Takami}, H., eds., {\em Society of Photo-Optical Instrumentation Engineers
  (SPIE) Conference Series} {\bf 10702},  107025L (July 2018).

\bibitem{Wright2018}
{Wright}, S.~A., {Horowitz}, P., {Maire}, J., {Werthimer}, D., {Antonio}, F.,
  {Aronson}, M., {Chaim-Weismann}, S., {Cosens}, M., {Drake}, F.~D., {Howard},
  A.~W., {Marcy}, G.~W., {Raffanti}, R., {Siemion}, A. P.~V., {Stone}, R.
  P.~S., {Treffers}, R.~R., and {Uttamchandani}, A., ``{Panoramic optical and
  near-infrared SETI instrument: overall specifications and science program},''
  in [{\em Ground-based and Airborne Instrumentation for Astronomy
  VII}{\nolinebreak\hspace{0.1em}]},  {Evans}, C.~J., {Simard}, L., and
  {Takami}, H., eds., {\em Society of Photo-Optical Instrumentation Engineers
  (SPIE) Conference Series} {\bf 10702},  107025I (July 2018).

\bibitem{Cosens2018}
{Cosens}, M., {Maire}, J., {Wright}, S.~A., {Antonio}, F., {Aronson}, M.,
  {Chaim-Weismann}, S.~A., {Drake}, F.~D., {Horowitz}, P., {Howard}, A.~W.,
  {Raffanti}, R., {Siemion}, A. P.~V., {Stone}, R. P.~S., {Treffers}, R.~R.,
  {Uttamchand ani}, A., and {Werthimer}, D., ``{Panoramic optical and
  near-infrared SETI instrument: prototype design and testing},'' in [{\em
  Ground-based and Airborne Instrumentation for Astronomy
  VII}{\nolinebreak\hspace{0.1em}]},  {Evans}, C.~J., {Simard}, L., and
  {Takami}, H., eds., {\em Society of Photo-Optical Instrumentation Engineers
  (SPIE) Conference Series} {\bf 10702},  107025H (July 2018).

\bibitem{Wright2019}
{Wright}, S., {Antonio}, F.~P., {Aronson}, M.~L., {Chaim-Weismann}, S.~A.,
  {Cosens}, M., {Drake}, F.~D., {Horowitz}, P., {Howard}, A.~W., {Liu}, W.,
  {Maire}, J., {Siemion}, A. P.~V., {Raffanti}, R., {Shippee}, G.~D., {Stone},
  R. P.~S., {Treffers}, R.~R., {Uttamchandani}, A., {Werthimer}, D., and
  {Wiley}, J., ``{Panoramic SETI: An all-sky fast time-domain observatory},''
  in [{\em Bulletin of the American Astronomical
  Society}{\nolinebreak\hspace{0.1em}]},   {\bf 51},  264 (Sept. 2019).

\bibitem{Cao2021a}
{Cao}, Z., {Aharonian}, F.~A., and {An, Q. et al.}, ``{Ultrahigh-energy photons
  up to 1.4 petaelectronvolts from 12 {\ensuremath{\gamma}}-ray Galactic
  sources},'' {\em \nat}~{\bf 594},  33--36 (June 2021).

\bibitem{2020NIMPA.95862113B}
{Budnev}, N., {Astapov}, I.~I., {Bezyazeekov}, P.~A., {Borodin}, A.,
  {Br{\"u}ckner}, M., {Chernykh}, D., {Chiavassa}, A., {Dyachok}, A.,
  {Fedorov}, O., {Gafarov}, A., {Garmash}, A., {Grebenyuk}, V., {Gress}, O.,
  {Gress}, T., {Grinyuk}, A., {Grishin}, O., {Horns}, D., {Ivanova}, A.,
  {Kalmykov}, N., {Kazarina}, Y., {Kindin}, V., {Kiryuhin}, S., {Kokoulin}, R.,
  {Komponiets}, K., {Korosteleva}, E., {Kostunin}, D., {Kozhin}, V.,
  {Kravchenko}, E., {Kryukov}, A., {Kuzmichev}, L., {Lagutin}, A.,
  {Lubsandorzhiev}, B., {Lubsandorzhiev}, N., {Mirgazov}, R., {Mirzoyan}, R.,
  {Monkhoev}, R., {Osipova}, E., {Pakhorukov}, A., {Pan}, A., {Panasyuk}, M.,
  {Pankov}, L., {Petrukhin}, A., {Poleschuk}, V., {Popesku}, M., {Popova}, E.,
  {Porelli}, A., {Postnikov}, E., {Prosin}, V., {Ptuskin}, V., {Pushnin}, A.,
  {Raikin}, R., {Rubtsov}, G., {Ryabov}, E., {Sagan}, Y., {Samoliga}, V.,
  {Silaev}, A., {Silaev}, A., {Sidorenkov}, A., {Skurikhin}, A., {Slunecka},
  M., {Sokolov}, A., {Spiering}, C., {Sveshnikova}, L., {Suvorkin}, Y.,
  {Tabolenko}, V., {Tarashchansky}, B., {Tkachev}, L., {Tluczykont}, M.,
  {Ushakov}, N., {Ustinov}, K., {Vaidyanathan}, A., {Voronin}, D.,
  {Wischnewski}, R., {Zagorodnikov}, A., {Zurbanov}, V., {Zhurov}, D., and
  {Yashin}, I., ``{TAIGA-A hybrid array for high-energy gamma astronomy and
  cosmic-ray physics},'' {\em Nuclear Instruments and Methods in Physics
  Research A}~{\bf 958},  162113 (Apr. 2020).

\bibitem{Moreira2009}
{Moreira}, P., {Serrano}, J., {Wlostowski}, T., {Loschmidt}, P., and {Gaderer},
  G., ``White rabbit: Sub-nanosecond timing distribution over ethernet,'' in
  [{\em 2009 International Symposium on Precision Clock Synchronization for
  Measurement, Control and Communication}{\nolinebreak\hspace{0.1em}]},   1--5
  (2009).

\bibitem{Liu2020}
{Liu, Wei et al}, ``Panoramic seti: Electronics, timing, and network design,''
  {\em Society of Photo-Optical Instrumentation Engineers (SPIE), this
  conference} (Dec. 2020).

\bibitem{Wright2001}
{Wright}, S.~A., {Drake}, F., {Stone}, R.~P., {Treffers}, D., and {Werthimer},
  D., ``{Improved optical SETI detector},'' in [{\em The Search for
  Extraterrestrial Intelligence (SETI) in the Optical Spectrum
  III}{\nolinebreak\hspace{0.1em}]},  {Kingsley}, S.~A. and {Bhathal}, R.,
  eds., {\em Society of Photo-Optical Instrumentation Engineers (SPIE)
  Conference Series} {\bf 4273},  173--177 (Aug. 2001).

\bibitem{Wright2014}
{Wright}, S.~A., {Werthimer}, D., {Treffers}, R.~R., {Maire}, J., {Marcy},
  G.~W., {Stone}, R. P.~S., {Drake}, F., {Meyer}, E., {Dorval}, P., and
  {Siemion}, A., ``{A near-infrared SETI experiment: instrument overview},'' in
  [{\em Ground-based and Airborne Instrumentation for Astronomy
  V}{\nolinebreak\hspace{0.1em}]},  {Ramsay}, S.~K., {McLean}, I.~S., and
  {Takami}, H., eds., {\em Society of Photo-Optical Instrumentation Engineers
  (SPIE) Conference Series} {\bf 9147},  91470J (July 2014).

\bibitem{Maire2020}
{Maire}, J., {Wright}, S.~A., {Werthimer}, D., {Antonio}, F.~P., {Brown}, A.,
  {Horowitz}, P., {Lee}, R., {Liu}, W., {Raffanti}, R., {Wiley}, J., {Cosens},
  M., {Heffner}, C.~M., {Howard}, A.~W., {Stone}, R. P.~S., and {Treffers},
  R.~R., ``{Panoramic SETI: on-sky results from prototype telescopes and
  instrumental design},'' in [{\em Society of Photo-Optical Instrumentation
  Engineers (SPIE) Conference Series}{\nolinebreak\hspace{0.1em}]},  {\em
  Society of Photo-Optical Instrumentation Engineers (SPIE) Conference Series}
  {\bf 11454},  114543C (Dec. 2020).

\bibitem{Weekes2002}
{Weekes}, T.~C., {Badran}, H., {Biller}, S.~D., {Bond}, I., {Bradbury}, S.,
  {Buckley}, J., {Carter-Lewis}, D., {Catanese}, M., {Criswell}, S., {Cui}, W.,
  {Dowkontt}, P., {Duke}, C., {Fegan}, D.~J., {Finley}, J., {Fortson}, L.,
  {Gaidos}, J., {Gillanders}, G.~H., {Grindlay}, J., {Hall}, T.~A., {Harris},
  K., {Hillas}, A.~M., {Kaaret}, P., {Kertzman}, M., {Kieda}, D., {Krennrich},
  F., {Lang}, M.~J., {LeBohec}, S., {Lessard}, R., {Lloyd-Evans}, J., {Knapp},
  J., {McKernan}, B., {McEnery}, J., {Moriarty}, P., {Muller}, D., {Ogden}, P.,
  {Ong}, R., {Petry}, D., {Quinn}, J., {Reay}, N.~W., {Reynolds}, P.~T.,
  {Rose}, J., {Salamon}, M., {Sembroski}, G., {Sidwell}, R., {Slane}, P.,
  {Stanton}, N., {Swordy}, S.~P., {Vassiliev}, V.~V., and {Wakely}, S.~P.,
  ``{VERITAS: the Very Energetic Radiation Imaging Telescope Array System},''
  {\em Astroparticle Physics}~{\bf 17},  221--243 (May 2002).

\bibitem{Fegan1997}
{Fegan}, D.~J., ``{TOPICAL REVIEW: $\gamma$/hadron separation at TeV
  energies},'' {\em Journal of Physics G Nuclear Physics}~{\bf 23},  1013--1060
  (Sept. 1997).

\end{thebibliography}
\bibliographystyle{spiebib} 

\end{document}